# Manipulation of photonic topological edge and corner states via trivial claddings

*Hai-Xiao Wang\*, Li Liang, Shuai Shao, Shiwei Tang, Junhui Hu, Yin Poo\* and Jian-Hua Jiang\**

Prof. H.-X. Wang, Prof. S. Tang
School of Physical Science and Technology
Ningbo University
Ningbo 315211, China
E-mail: wanghaixiao@nbu.edu.cn

L. Liang, Prof. Y. Poo
School of Electronic Science and Engineering
Nanjing University
Nanjing 210093, China
E-mail: ypoo@nju.edu.cn

S. Shao, Prof. J. Hu
School of Physical Science and Technology
Guangxi Normal University
Guilin 541004, China

Prof. J.-H Jiang
School of Physical Science and Technology, & Collaborative Innovation Center of Suzhou Nano Science and Technology
Soochow University
Suzhou 215006, China
Suzhou Institute for Advanced Research
University of Science and Technology of China
Suzhou 215123, China
E-mail: jhjiang3@ustc.edu.cn

**Keywords:** Topological crystalline insulator, kagome photonic crystal, topological edge and corner states

**Crystalline symmetry offers a powerful tool to realize photonic topological phases, in which additional trivial claddings are typically required to confine topological boundary**

states. However, the utility of the trivial cladding in manipulating topological waves is often overlooked. Here, we demonstrate two topologically distinct kagome photonic crystals (KPCs) based on different crystalline symmetries: $C_6$- symmetric KPCs exhibit a quantum spin Hall phase, while $C_3$-symmetric KPCs serve as trivial cladding. By tuning the geometric parameter of the trivial cladding, we observe that a pair of topological interface states featured with pseudospin-momentum locking undergoes a phase transition, accompanied by the appearance and disappearance of corner states in a finite hexagonal supercell. Such a geometry-induced band inversion is characterized by a sign change in the Dirac mass of the topological interface states and holds potential for applications such as rainbow trapping. Furthermore, we experimentally demonstrate the corner states, which is a hallmark of higher-order topology, also depend critically on the trivial cladding. Our work highlights the crucial role of trivial claddings on the formation of topological boundary states, and offers a novel approach for their manipulation.

## 1. Introduction

The fundamental interplay between topology and symmetry has revolutionized the classification of the phase of matter[1–5]. A representative paradigm is topological crystalline insulators (TCIs), which exploit crystalline symmetry to achieve topological states analogous to those topological insulators in materials without spin-orbital coupling[6]. Benefiting from great diversity of symmetry group, crystalline symmetry offers unprecedented opportunities for realizing exotic topological phases in photonics[7–16]. A landmark of photonic TCIs was the proposal of all-dielectric topological insulators based on expanded and shrunken honeycomb photonic crystals[7], where pseudospin degree of freedom are synthesized via crystalline symmetry combined with time-reversal symmetry. With the progressive understanding of TCIs, it has become clear that when the symmetry of the edge boundaries is lower than the bulk, predicted topological edge states often become gapped. These gapped edge states can themselves constitute lower-dimensional topological insulator, leading to the discovery of higher-order topology[4,17,18].

Generally, trivial claddings are essential for observing topological boundary states in photonic TCIs, primarily because they can prevent the topological boundary states, of which the frequencies are above light cone, from leaking into free space. However, unlike the rigorous bulk-edge correspondence in conventional topological insulators, the emergence of corner states also rely on the boundary configurations. For example, it is reported that multiple corner states are observed in $C_3$-symmetric photonic crystals with different outer claddings, some of which originate from nearest neighboring coupling, rather than the higher-order band topology[19–21]. Recent studies suggest that topological defects, such as dislocation[22–24], disclination[25,26], provide powerful probes for higher-order topology. While the emergence of topological boundary states is commonly attributed to the bulk topology, it has recently been recognized that the dispersion of the topological edge states can be effectively tuned by modifying the boundary configurations of the bulk structures[27,28]. Nevertheless, the utility of trivial claddings in manipulating topological waves and corner states is often overlooked.

Here, we demonstrate that the trivial claddings play a vital role in topological wave manipulations by systematically study kagome photonic crystals (KPCs). We first identify $C_6$-symmetric KPCs as a photonic analog of quantum spin Hall insulator, featuring pseudospin-dependent interface states. Using $C_3$-symmetric KPCs as trivial claddings, we show that the dispersion of the pseudospin-dependent interface states can be effectively tuned by adjusting the geometric parameter of the trivial cladding, enabling potential applications such as rainbow trapping. Moreover, corner states, which typically regarded as hallmarks of higher-order topology, can merge into or emerge from the bulk states simply by modifying the outer cladding. Our work underscores the utility of trivial claddings in forming topological boundary states, and suggests an alternative approach to manipulate topological boundary states.

## 2. Results and Discussion

In Chemistry, isomers are compounds with identical formulas but different atomic arrangements as well as distinct properties. Transferring this concept to topological physics, we consider two-dimensional kagome photonic crystals (KPCs) with three different primitive cell configurations according to their different center positions [see Fig. 1(a)]. For convenience, we termed the $C_3$-symmetric primitive cells centered at Maximal Wyckoff positions "$1b$" and "$1c$" as U-KPCs and D-KPCs, respectively, and the $C_6$-symmetric primitive cell centered at Maximal Wyckoff position "$1a$" as H-KPCs. The dielectric rods have radius $r = 0.1a$, where $a$ refers to the lattice constant, and the primitivity $\varepsilon = 11.7$. For U-KPCs and D-KPCs, the distance between the center of rod and primitive cell is defined as $d$. Throughout this work, we only consider the transverse-magnetic (TM) modes, and all eigen calculations are carried out with the radio-frequency module of the commercial software COMSOL MULTIPHYSICS. Figure 1(b) depicts the photonic band structure of the KPCs, where a complete band gap between the third and fourth band is observed. In what follows, we focus on the topological properties of this band gap. Obviously, the three KPCs with different primitives cell choices have the same band structures. Nevertheless, it should be emphasized that they are of different band topologies due to the distinct crystalline symmetry.

Following Ref. [29,30], we use the symmetry indicators to characterize the band topology of KPCs with different primitive cells choices. For $C_6$- and $C_3$-symmetric TCIs, topological indices are $\chi^{(6)} = \left(\left[M_1^{(2)}\right], \left[K_1^{(3)}\right]\right)$ and $\chi^{(3)} = \left(\left[K_1^{(3)}\right], \left[K_2^{(3)}\right]\right)$, respectively, where $\left[\Pi_p^{(n)}\right] = \#\Pi_p^{(n)} - \#\Gamma_p^{(n)}$ is an integer invariants, and $\#\Pi_p^{(n)}$ is the number of states in the within the frequency bands having rotation eigenvalues $\Pi_p^{(n)} = e^{\frac{2\pi i (p-1)}{n}}$ ($p = 1,2,\dots,n$). By checking the symmetry eigenvalue at high symmetry points, the $C_6$- and $C_3$-symmetric topological indices of H-KPCs are $\chi^{(6)} = (-2,0)$ and $\chi^{(3)} = (0,1)$, indicating it is of nontrivial topology (see details in Supporting Information). Since $C_6$-symmetric topological indices are undefined for U-KPCs and D-KPCs, we classify them as trivial insulator in sense of $C_6$-symmetric band topology. Moreover, it is deserved to notice that the $C_6$ point group has two doubly degenerate representations, which are isomorphic to the $p$ and $d$ modes at Γ point[7,31,32] [see Fig. 1(c)]. These modes can be further reorganized into a pair of modes with orbital angular momentum, namely $p_\pm = p_x \pm i p_y$ and $d_\pm = d_{x^2-y^2} \pm i d_{xy}$, in which the positive (negative) sign mimics the spin-up (spin-down) states. In addition, the $p$ doublet and $d$ doublet have opposite parities, and the $d$ doublet has a higher frequency than that of $p$ doublet. As shown in Refs. [31], the $k \cdot p$ Hamiltonian around Γ point in $C_6$-symmetric photonic crystals is an analog of the quantum-spin Hall effect in electronics and the $p$-$d$ inversion at the Γ point leads to the formation of photonic topological insulator.

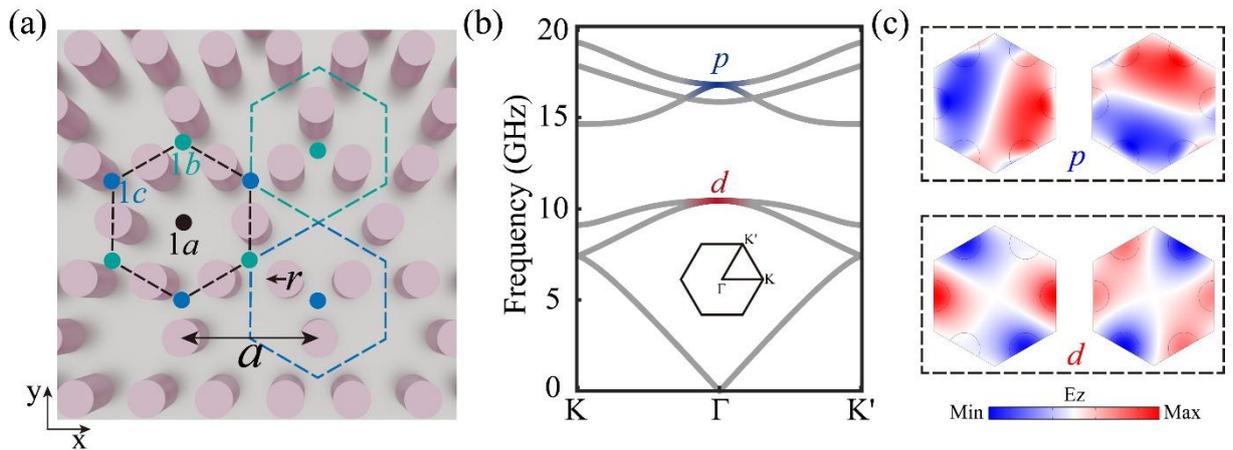

**Figure 1** | (a) Schematic of topological isomers based on KPCs, where three primitive cell configurations, i.e., H-KPCs, U-KPCs, and D-KPCs, centered at different Wycoff positions are presented. (b) The band structure of KPCs. Note that the parities of states at Γ point are well

defined only for H-KPCs. (c) The electric field patterns of two pairs of degenerated modes featured with odd (upper panel) and even (lower panel) parities at Γ point.

According to the bulk-boundary correspondence, it is predicted that a pair of helical edge states emerge at the boundary of H-KPCs. Although the free space can be regarded as trivial media, the edge states may leaky into air since the frequency of emergent edge states may above the light cone. To this end, we take the D-KPCs as the trivial cladding [see Fig. 2(a)] since that the H-KPCs and D-KPCs have identical band structures and therefore conveniently overlapping bandgaps while having different topological indices. Figure 2(b) depicts the simulated and measured interface states, where a gap originated from the symmetry breaking at the boundary is observed. The dispersions of the topological interface states are measured in our experiments using the Fourier-transformed electromagnetic field scanning method and are in good agreement with the simulated results. To visualize the spin Hall nature of the system, we present the phase profiles of the field pattern as well as the Poynting vectors of the topological interface states at a finite wavevector in the lower panel in Fig. 2(c). It is seen that there are phase singularity points, i.e., the phase vortices, around which the phase winds $\pm 2\pi$, indicating finite orbital angular momenta. The phase vortices have opposite winding directions for opposite wavevectors, indicating opposite angular momenta for opposite wavevectors. The pseudospin-momentum locking effect is also demonstrated by checking the Poynting vector of two edge states with opposite momenta.

In addition, the mirror symmetry guarantees the topological interface states at $k_x = 0$ host either even or odd parity [see upper panel in Fig. 2(c)]. Hence, the gapped interface states can be further described by the 1D massive Dirac equations, resembling the time-reversal symmetry broken edge states in quantum spin Hall insulators when the edges develop magnetic orders due to interactions or magnetic doping[33,34]. We then employ Dirac mass $m$ to quantify the frequency difference between the odd- and even-parity states[35,36], i.e., $\omega_+ - \omega_-$ , where the subscripts "$\pm$" represent the upper and lower topological interface states at $k_x = 0$. Intriguingly, the frequency gap size depends critically on the D-KPCs. As shown in Fig. 2(d), by continuously tuning the geometric parameter $d$ of the trivial D-KPCs, the frequency gap of the topological interface states gradually decrease to 0 at $d = 0.675l$, and reopens as $d$ increases further, indicating a topological phase transition occurs. Although this phase transition is fine-tuning, which also can be achieved by alternating the dielectric permittivity of the outer cladding or varying the geometry of the edge unit cells, it is should be emphasized that the out cladding is essential to manifest the pseudospin dependent topological interface states and the topological phase transition of the interface states [see details in Supporting Information]. In Fig. 2(e), we further present the calculated eigen spectra and the measured interface states when $d = 0.75l$. It is seen that the topological interface states are separated by a small gap, which the topological interface state with even parity has a higher frequency than that with odd parity [see the eigen electric field patterns in the upper panel of Fig. 2(f)]. In experiment, the Fourier spectrum of the interface states around $k_x = 0$ has a larger amplitude than others owing to the dramatical change of the interface states at $k_x = 0$. Besides, the interface states still hold for the pseudospin-momentum locking, which can be inferred by examining the phase profiles of the field pattern as well as the Poynting vectors of the topological interface states at a finite wavevector away from the time-reversal invariant point [see the lower panel of Fig. 2(f)].

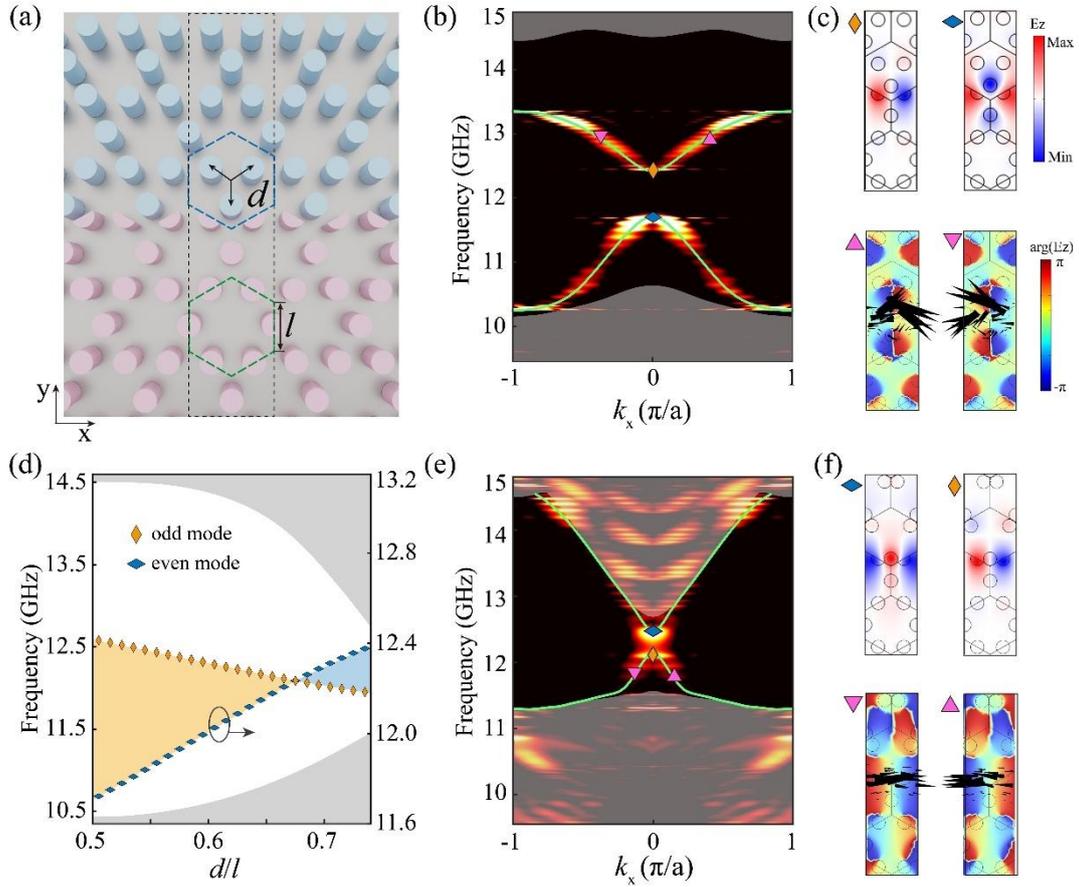

**Figure 2** | (a) Schematic of domain wall formed by H-KPCs and D-KPCs with different geometric parameter $d$. (b) The simulated and experimental edge dispersions of the domain wall formed by H-KPCs and D-KPCs with $d = 0.5l$. (c) Upper panel: electric field patterns of edge states at $k_x = 0$. Lower panel: phase distribution and Poynting vectors of edge states at $k_x = 0.1\pi/a$. (d) Evolution of the frequency gap versus the geometric parameter $d$. The gray areas refer to bulk states. (e) The simulated and experimental edge dispersions of the domain wall formed by H-KPCs and D-KPCs with $d = 0.75l$. (f) Upper panel: electric field patterns

of edge states at $k_x = 0$. Lower panel: phase distribution and Poynting vectors of edge states at $k_x = 0.05\pi/a$.

Thanks to the tunability of the topological interface states, it may find potential application on topological rainbow trapping. As illustrated in Fig. 3(a), we design a graded structure in which the geometric parameter $d$ of the D-KPCs varies from 0.35 to 0.56 along $x$-direction while keep H-KPCs remains unchanged. Such gradual variations in lattice geometry reshape the dispersion of the topological interface states, progressively reducing the group velocity for different frequency components and enabling their spatial separation. Here, we present the evolution of the upper branch of topological interface states and the corresponding group velocities versus under different geometric parameters in Fig. 3(b). It is seen that the both the dispersion and the corresponding group velocity varies when the geometric parameter $d$ changes, which provides a degree of freedom to modulate the propagating of the topological edge states along the interface. As shown in Fig. 3(c), when a beam of light in the frequency range of the topological interface on the left side is input, the electromagnetic wave with different frequencies can be separated and trapped at different positions. To verify it, we implement a microwave experiment by placing an excited point source on the left side of the graded structure with different frequencies. As shown in Fig. 3(d), different frequencies with different intensity distributions in spatial positions are clearly observed, identifying the topological rainbow trapping effect in our graded structure. We remark that the lower branch of the topological interface state also exhibits similar tunability and holds potential application for topological rainbow trapping [see details in Supporting Information]. Unlike the previous studies that used a single interface band for topological rainbow trapping, our design effectively extends the working frequency of topological rainbow trapping by taking advantage of two tunable interface bands.

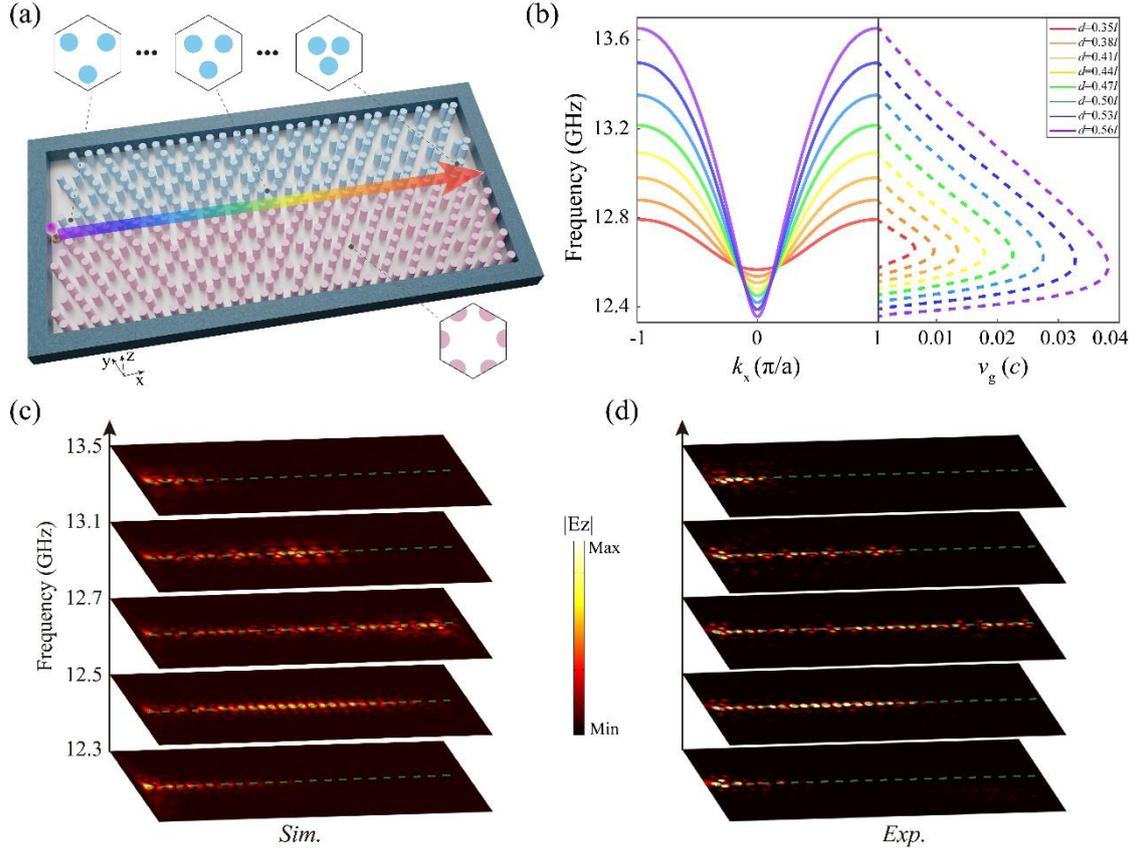

**Figure 3** | (a) Photo of topological rainbow trapping in the interface between H-KPCs and D-KPCs. (b) The dispersion (left panel) and corresponding group velocity (right panel) of the upper branch of topological interface states versus different geometric parameter. (c) Simulated and (d) experimental normalized electric intensity distribution of $E_z$ under different frequencies.

Furthermore, the tunability of the topological interfaces states also provides a playground to study the higher-order topology. As an illustration, we construct a finite hexagon-shaped supercell, where H-KPCs are surrounded by D-KPCs [see Fig. 4(a)]. Generally, the trivial cladding is also essential to manifest corner states in the edge gap. Note that the hexagon-shaped supercell does not exhibit $C_6$ symmetry owing to the $C_3$- symmetric nature of outer claddings. Figure 4(b) presents the eigen spectra of the hexagon-shaped supercell, in which six corner states within the bandgap are observed. Remarkably, the emerged corner states are divided into two sets because of the finite size effect [see details in Supporting Information], each of which compose of three degenerated corner states owing to the $C_3$-symmetry nature of the hexagon-shaped supercell. Figure 3(c) depicts the two representative electric field patterns of the corner states. It is seen that the eigenstate with higher (lower) frequency indicated by blue triangular (red star) mainly localized at the lower two corners, and is even (odd) symmetric with respect to their midplanes. This mainly attributes to the coupling between two neighboring corner states. Moreover, we calculate the corner charge of the topological core and trivial cladding, which gives $Q_{inner} = 0.5$ and $Q_{outer} = 0$, respectively, indicating that the corner states are originated from the corner charge mismatch between topological core and trivial cladding.

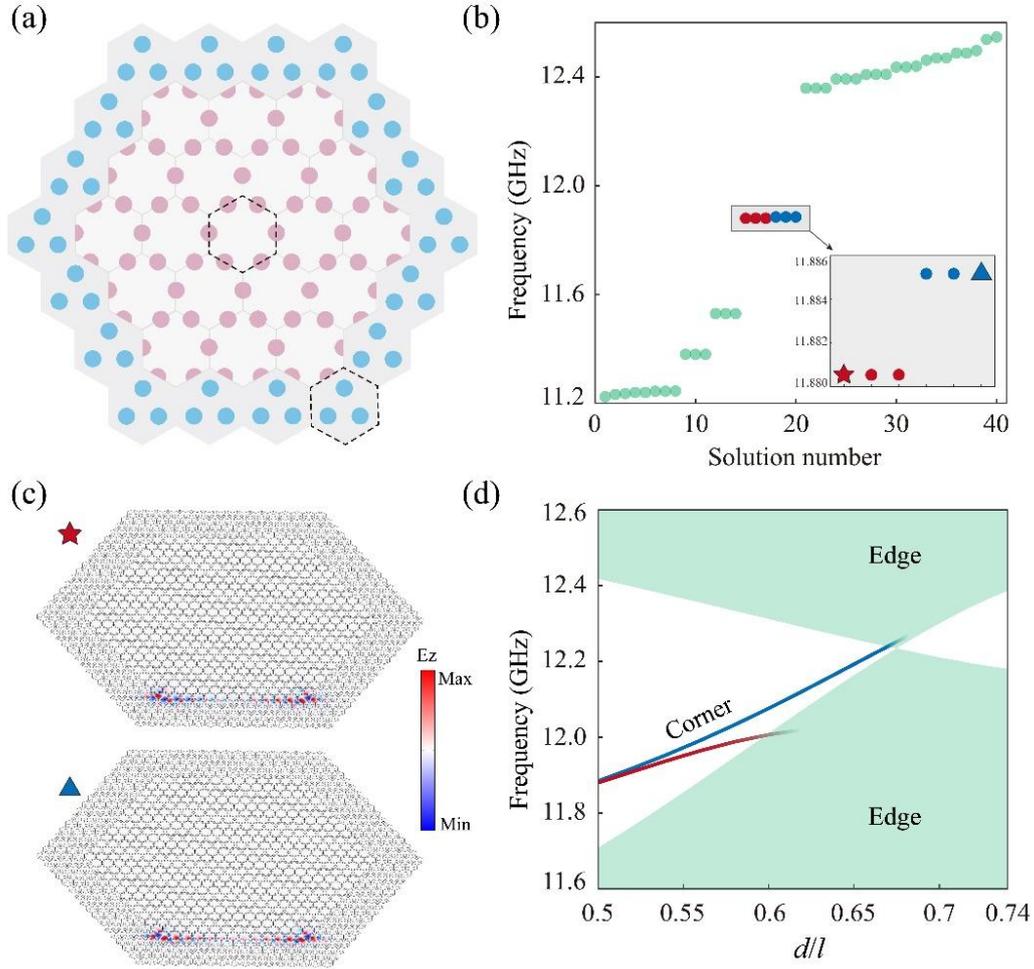

**Figure 4** | (a) Schematic of $C_3$ symmetric supercell, where H-KPCs are surrounded by U-KPCs. (b) Eigen spectra of finite supercell. The edge and corner states are colored in green and bule (red), respectively. (c) Two typical electric fields of corner states. (d) Evolutions of corner states versus the geometric parameter $d$.

We proceed to study the tunability of the corner states by gradually changing the geometric parameter $d$ of trivial claddings (D-KPCs) while keep the topological core (H-KPCs) unchanged. As shown in Fig. 4(d), the corner states gradually emerged into the edge states accompany with the closing of the gap of the topological interface states. In particular, the corner states no longer appear at the reopening gaps although the higher-order band topology and the corner charge of the H-KPCs remains unchanged. This evidence that the corner states can be effectively tuned by the boundary configurations. To further unveil the appearance and disappearance of the corner states, we pay special attention to the interface configuration of the hexagon-shaped supercell by considering the corresponding tight-binding model. The analytical results show that tuning the geometric parameter is equivalent to change the ratio between intercell and intracell hoppings in a Su-Schrieffer-Heeger (SSH) model [see details in Supporting Information]. The larger (smaller) geometric parameter indicates a stronger intracell (intercell) hopping, resulting in a trivial (nontrivial) SSH model with the disappearance (appearance) of the end states. From this perspective, the corner states can be regarded as the physical manifestation of nontrivial topological phase of the interface configuration in the hexagon-shaped supercell.

To visualize the bulk, edge and corner states in spectral measurements, we implement three types of pump-probe measurement for detecting the bulk, edge and corner mode. Specifically, we use the half-hexagonal structure to avoid whisper-gallery modes in a closed boundary as shown in Fig. 5(a). Transmission spectra for the three different pump-probe measurements are presented in Fig. 5(b), clearly indicating the edge and corner modes emerging in the bulk bandgap. The gap of the topological interface band is indicated by two dashed lines. To excite the corner states with even (odd) parity, we employ two point-sources with same (opposite) phases. For further confirmation of the localization of corner states, we map out the out-of-field scanning method and find that the field is exponentially localized at both left and right corners. The measured and simulated electric field profiles for both even- and odd-parity corner states are displayed in Figs. 5(c) and 5(d), respectively, indicating good agreement with each other. Finally, we point out that the corner states in our work do not exhibit with the pseudospin dependent nature owing to the absence of $C_6$ symmetry, in strong contrast to the corner states in the higher-order quantum spin Hall insulators [18].

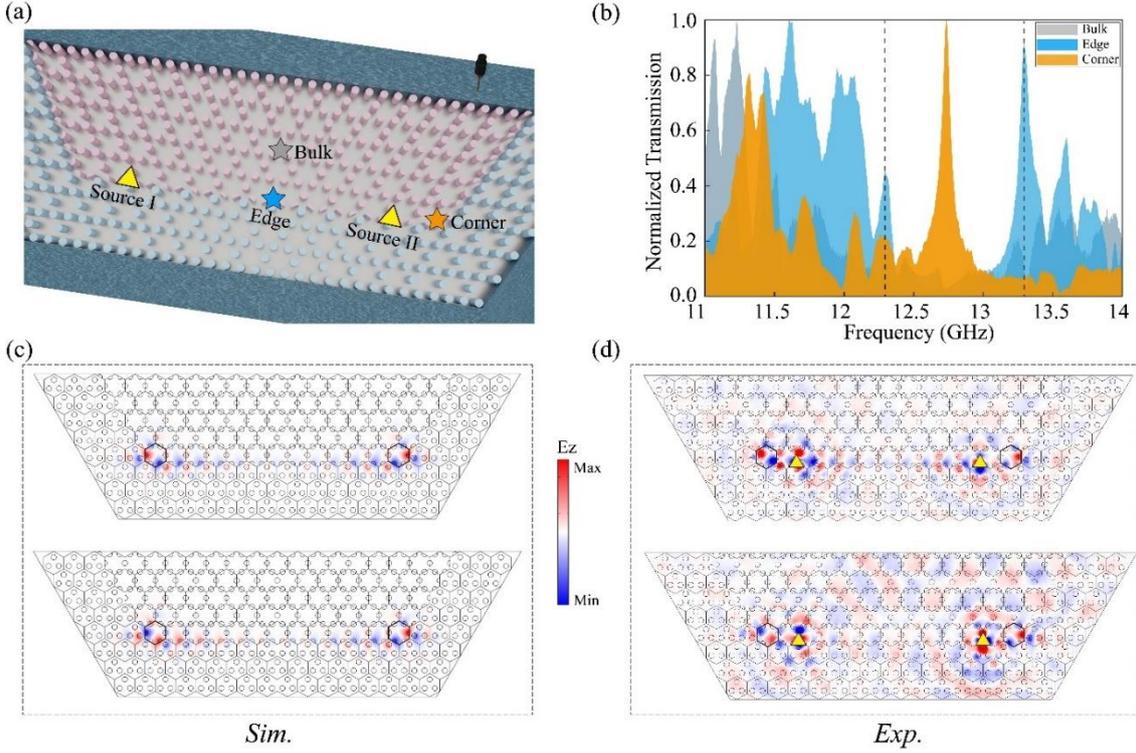

**Figure 5** | (a) Schematic of the half-hexagonal structure composed of H-KPCs (inner) and U-KPCs (outer cladding) to visualize the corner states. (b) Measured transmission spectra for three types of pump-probe: bulk (gray), edge (blue), and corner (orange) pump-probe measurements. (c, d) Electric field patterns for the corner states from (c) simulations and (d) experiment, which are in good agreement with each other.

## 3. Conclusion

In conclusion, we demonstrate that the trivial claddings play a vital role in topological wave manipulations. Using KPCs with different crystalline symmetries, we first identify that the $C_6$-symmetric KPCs is of quantum spin Hall phase, supporting the pseudospin-dependent topological interface states. By employing $C_3$-symmetric KPCs as trivial cladding, we find the dispersion of helical interface states can be effectively tuned by adjusting the geometric parameter of the trivial cladding, which may find potential application on topological rainbow trapping. Moreover, corner states, which commonly regarded as a hallmark of higher-order topology, can be merged into or emerge from the bulk states simply by modifying the outer cladding. Our findings reveal claddings as indispensable design parameters for reconfigurable topological photonic systems, and provide an alternative approach to manipulate topological boundary states.

**Acknowledgements**

The authors thank the support from Natural Science Foundation of Guangxi Province (Grant No. 2023GXNSFAA026048), the start-up funding of Ningbo University, National Key Research and Development Program of China (Grant No. 2022YFA1404400), the "Hundred Talents Program" of the Chinese Academy of Sciences, and National Natural Science Foundation of China (Grants No. 12474432 No. 11904060, No. 12074281, and No. 12125504), and the Priority Academic Program Development of Jiangsu Higher Education Institutions.

H.-X.W and L.L contributed equally to this work.

**Conflict of Interest**

The authors declare no conflict of interest.

**Data Availability Statement**

The data that support the findings of this study are available from the corresponding author upon reasonable request.